\documentclass{aa}
\usepackage{graphicx}
\usepackage{txfonts}
\usepackage[authoryear]{natbib}
\bibpunct{(}{)}{;}{a}{}{,}

\newcommand{\sax} {\object{SAX J2103.5+4545}}
\newcommand{\xte} {\object{XTE J1858+034}}
\newcommand{\gro} {\object{GRO J2058+42}}
\newcommand{\igr} {\object{IGR J01363+6610}}
\newcommand{\igrj} {\object{IGR J00370+6122}}

\def\simless{\mathbin{\lower 3pt\hbox
     {$\rlap{\raise 5pt\hbox{$\char'074$}}\mathchar"7218$}}}   
\def\simmore{\mathbin{\lower 3pt\hbox
     {$\rlap{\raise 5pt\hbox{$\char'076$}}\mathchar"7218$}}}   

\def\msun{~{\rm M}_\odot}
\def\rsun{~{\rm R}_\odot}

\begin{document}

\title{Identification of the optical counterparts of high-mass X-ray
binaries through optical photometry and spectroscopy}

\subtitle{}

\author{P. Reig\inst{1,2}
\and I. Negueruela\inst{3}
\and G. Papamastorakis\inst{1,2}
\and A. Manousakis\inst{2}
\and T. Kougentakis\inst{1}
}

\institute{
IESL, Foundation for Research and Technology, 711 10 Heraklion, Crete,
Greece\\ 
\and University of Crete, Physics Department, PO Box 2208, 710 03
Heraklion, Crete, Greece \\
\and Departamento de F\'{\i}sica, Ingenier\'{\i}a de Sistemas y Teor\'{\i}a
de la Se\~nal, Universidad de Alicante, E-03080 Alicante, spain\\
}

\authorrunning{Reig et~al.}
\titlerunning{Identification of new Be/X-ray binaries}

\offprints{P. Reig, \\ \email{pau@physics.uoc.gr}}

\date{Received / Accepted }

\abstract{We present the results of our search for optical counterparts to
high-mass X-ray transient sources discovered by various X-ray missions. We
obtained CCD images of the X-ray fields through $BVR$ and H$\alpha$
filters in order to identify early-type stars in the $R-H\alpha$ versus
$B-V$ colour-colour diagram. We also obtained medium-resolution
spectroscopy of the candidates in order to confirm the presence of
H$\alpha$ emission and perform spectral classification. We report on the
discovery of the optical counterparts to two X-ray sources: \xte\ and
\igr, and the follow-up observations of another two, newly identified by
our group: \sax\ and \gro. For another source, \igrj, we present the first
detailed optical spectral analysis. The optical photometry and
spectroscopy reveal B-type companions in all five sources; \gro, \sax\ and
\igr\ are positively identified with Be/X-ray binaries, \igrj\
with a supergiant X-ray binary, while the nature of 
\xte\ is uncertain. We also study the relationship between the optical
and X-ray emission during quiescent states.

}

\maketitle

\begin{figure*}
\begin{center}
\begin{tabular}{cc}
\includegraphics[width=8cm,height=6cm]{./xte1858_cd.eps} &
\includegraphics[width=8cm,height=6cm]{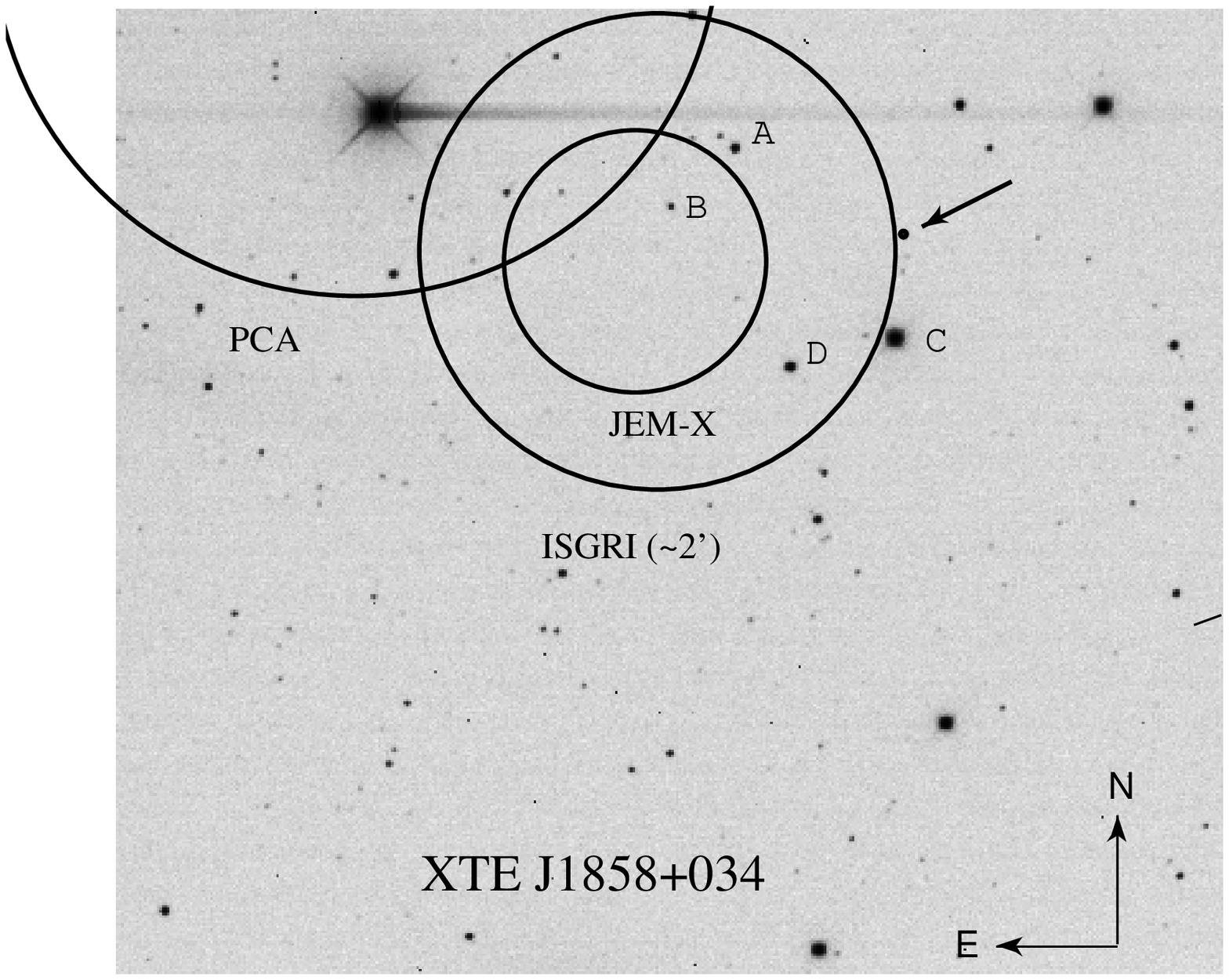} \\
\includegraphics[width=8cm,height=6cm]{./igr0136_cd.eps} &
\includegraphics[width=8cm,height=6cm]{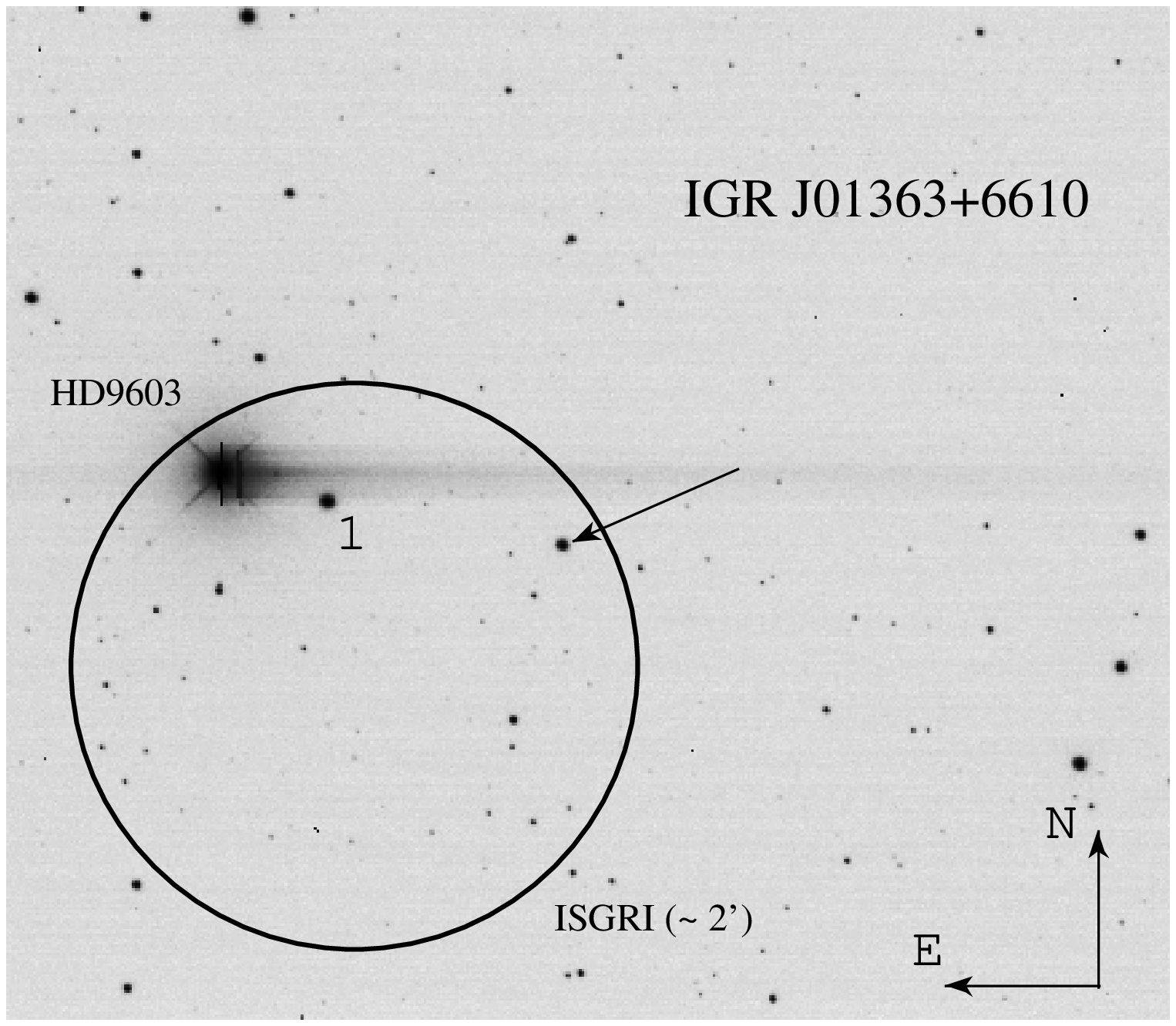} \\
\end{tabular}
\caption[]{Colour-colour diagrams and V-band images of the field around
\xte\ (top) and \igr\ (bottom). The images were taken from the 1.3m telescope 
of the Skinakas observatory on 20 May 2004 and 27 July 2004, respectively
and the diagrams used data obtained on 28 and 29 June 2004, respectively.
Other potential (and subsequently rejected) candidates are also indicated.
For \xte, the 2.5$^{\prime}$ PCA RXTE, 2$^{\prime}$ ISGRI
INTEGRAL and 1$^{\prime}$ JEM-X INTEGRAL radius error circles are shown.
For \igr, the 2$^{\prime}$ ISGRI INTEGRAL error circle is shown.
The proposed optical counterparts are marked with a filled circle on the
diagrams and with an arrow on the images. }
\label{cdimag}
\end{center}
\end{figure*}

\section{Introduction} \label{introduction}

The interest in X/$\gamma$-ray Astronomy has grown enormously in the last
decades thanks to the ability to send X-ray space missions above the
Earth's atmosphere. There are more than half a million X-ray sources
detected and over a hundred missions (past and currently operational)
devoted to the study of X/$\gamma$ rays.  With the improved sensibilities
of the currently operational missions new detections occur very
frequently.  Among these, X-ray binaries form an important group because
they contain neutron stars and black holes allowing the study of the
physics of matter under strong gravity and the physics of accretion.

X-ray binaries consist of a compact object orbiting a companion star. They
admit several classification schemes depending upon whether the emphasis
is put on their optical or X-ray properties. This work focuses on the
high-mass X-ray binaries (HMXB), that is,  systems formed by a magnetised
neutron star and a OB-type star  \citep[see][for a recent review]{neg05}. 
According to the luminosity class of the primary they further divide into
supergiant X-ray binaries (SXRB) and Be/X-ray binaries (BeX).  All but
four HMXBs harbour X-ray pulsars. By far, the most numerous group (about
70\%) is that of BeX \citep{coe00,zio02,neg04}.

The generally accepted definition of a Be star is that of a non-supergiant
B-type star whose spectrum shows or, has at some time shown, Balmer lines
in emission. An infrared excess when compared with non line-emitting B
stars of the same spectral type is another observational characteristic of
Be stars. The origin of these observational characteristics lies in a
quasi-Keplerian circumstellar disk around the equator of the Be star. This
disk is fed from material expelled from the rapidly rotating Be star in a
manner that it is not yet understood \citep{por03}. Recombination
radiation from ionised hydrogen in this hot ($T_{\rm disk}=0.5-0.8 T_{\rm
eff}$) extended envelope produces the Balmer line emission and the
infrared excess. Be stars may appear as isolated objects or taking part in
binaries. While isolated Be stars have been observed to sample all B type
subclasses, the spectral type of Be stars in galactic and LMC X-ray
binaries is confined to the narrow range O9-B2 \citep{neg98}. In the SMC,
however, the spectral distribution of Be in BeX shows a much greater
agreement with those of isolated Be stars \citep{coe05}.

In HMXBs, X/$\gamma$-rays are produced as the result of accretion of
matter onto the neutron star. The way how matter is transferred from the
massive star differs in each case. Among HMXBs with evolved companions
there are disk-fed and wind-fed systems depending upon whether accreted
matter is transferred through an accretion disk (with the mass donor star
being close to filling its Roche lobe) or via the stellar wind of the
evolved companion. In BeX, the reservoir of matter is provided by the
circumstellar disk.  Wind-fed SXRBs are persistent long-pulsing ($>
10^2$~s) X-ray sources with luminosities of the order of
$10^{35}-10^{36}$~erg~s$^{-1}$. Occasionally they exhibit flaring
variability on short time scales (seconds). With only one representative
in the Galaxy (Cen X-3) and three in total (SMC X-1 and LMC X-4), disk-fed
HMXBs display high X-ray luminosities ($L_{\rm X}\sim
10^{38}$~erg~s$^{-1}$) and short spin periods (a few seconds). The vast
majority of BeX are transient sources showing either regular orbital
modulated outbursts with  $L_{\rm X}=10^{35}-10^{36}$ erg s$^{-1}$, or
giant, unexpected outbursts with $L_{\rm X} \simmore 10^{37}$ erg
s$^{-1}$, or both. Persistent BeX also exist \citep{reig99}. Transient BeX
tend to contain fast rotating neutron star, while persistent BeX show pulse
periods above $\sim 200$ s \citep{neg05}.

A search in the {\it SIMBAD} database yields that only about one
third of HMXBs in the Milky Way have well-established optical
counterparts.  One of the largest hindrances to the identification of
optical counterparts is the size of the uncertainty in X-ray positions.
This can be as small as a few arcseconds for imaging detectors ({\it ROSAT
HRI}, {\it Chandra ACIS}, {\it XMM-Newton EPIC}), or as large as a few
degrees for all sky monitors ({\it CGRO BATSE}).  In addition, many
identifications are based only upon positional coincidence. In regions
with a high density of early type stars a danger exists that these
identifications may be spurious. Furthermore, given the transient nature
of these systems, many times only one detection of the source is
available.

An optical identification is necessary to facilitate a complete study of
these systems. Without a known counterpart, observations are limited to
X-ray energies, and hence our understanding of the structure and dynamics
of those systems that remain optically unidentified is incomplete. In this
work we provide optical information of five HMXBs and compare the optical
and X-ray properties during quiescence.

\begin{table*}
\begin{center}
\caption{Results of the photometric observations}
\label{phot}
\begin{tabular}{cccccc}
\noalign{\smallskip} \hline \hline
Date	&MJD	&B	&V	&R 	&I\\
\hline 
\multicolumn{6}{c}{\xte} \\
\hline
20-05-04 &53146.45 &19.63$\pm$0.01 &18.01$\pm$0.01 &16.94$\pm$0.01 &- \\
27-07-04 &53214.38 &19.60$\pm$0.02 &17.97$\pm$0.03 &16.92$\pm$0.04 &15.38$\pm$0.04 \\
24-08-04 &53242.41 &19.61$\pm$0.02 &18.00$\pm$0.02 &16.95$\pm$0.01 &15.32$\pm$0.02 \\
\hline
\multicolumn{6}{c}{\gro} \\
\hline
07-06-03$^a$ &52798.50 &16.03$\pm$0.03 &14.93$\pm$0.03 &14.23$\pm$0.03 &- \\
08-06-03$^a$ &52799.50 &16.07$\pm$0.03 &14.94$\pm$0.02 &14.24$\pm$0.02 &- \\
24-08-03$^a$ &52876.38 &16.03$\pm$0.02 &14.91$\pm$0.03 &14.26$\pm$0.03 &13.50$\pm$0.04 \\
05-07-04$^a$ &53192.44 &16.03$\pm$0.02 &14.89$\pm$0.02 &14.16$\pm$0.02 &13.35$\pm$0.02 \\
27-07-04$^a$ &53214.51 &16.07$\pm$0.02 &14.92$\pm$0.03 &14.19$\pm$0.03 &13.41$\pm$0.04 \\
24-08-04$^a$ &53243.52 &16.05$\pm$0.02 &14.96$\pm$0.02 &14.24$\pm$0.02 &13.46$\pm$0.02 \\
14-09-04    &53263.35 &16.04$\pm$0.02 &14.92$\pm$0.02 &14.22$\pm$0.02 &13.47$\pm$0.02 \\
01-10-04    &53280.30 &16.07$\pm$0.02 &14.96$\pm$0.02 &14.24$\pm$0.02 &13.47$\pm$0.02 \\
\hline
\multicolumn{6}{c}{\sax} \\
\hline
08-06-03$^b$ &52799.47 &15.34$\pm$0.02 &14.21$\pm$0.02 &13.48$\pm$0.02 &- \\
24-08-03$^b$ &52876.41 &15.35$\pm$0.02 &14.25$\pm$0.03 &13.57$\pm$0.03 &12.85$\pm$0.03 \\
20-05-04    &53146.56 &15.40$\pm$0.01 &14.32$\pm$0.01 &13.63$\pm$0.01 &12.87$\pm$0.01 \\
05-07-04    &53192.37 &15.42$\pm$0.02 &14.33$\pm$0.02 &13.64$\pm$0.02 &12.92$\pm$0.02 \\
24-08-04    &53243.48 &15.39$\pm$0.02 &14.31$\pm$0.02 &13.61$\pm$0.01 &12.85$\pm$0.02 \\
14-09-04    &53263.38 &15.39$\pm$0.02 &14.27$\pm$0.02 &13.58$\pm$0.02 &12.84$\pm$0.03 \\
01-10-04    &53280.37 &15.40$\pm$0.02 &14.30$\pm$0.02 &13.61$\pm$0.02 &12.85$\pm$0.02 \\
\hline
\multicolumn{6}{c}{\igrj} \\
\hline
20-05-04 &53146.57 &10.21$\pm$0.01 &9.65$\pm$0.01 &9.36$\pm$0.01 &8.88$\pm$0.01 \\
14-09-04 &53263.52 &10.24$\pm$0.02 &9.70$\pm$0.02 &9.36$\pm$0.02 &8.91$\pm$0.03 \\
\hline
\multicolumn{6}{c}{\igr} \\
\hline
27-07-04 &53214.55 &14.68$\pm$0.02 &13.29$\pm$0.03 &12.32$\pm$0.04 &11.37$\pm$0.04 \\
14-09-04 &53263.44 &14.70$\pm$0.02 &13.33$\pm$0.02 &12.35$\pm$0.02 &11.37$\pm$0.03 \\
01-10-04 &53280.40 &14.62$\pm$0.02 &13.26$\pm$0.02 &12.28$\pm$0.02 &11.31$\pm$0.02 \\
\noalign{\smallskip} \hline \hline
\multicolumn{6}{l}{$a$: From \citet{wil05}} \\
\multicolumn{5}{l}{$b$: From \citet{reig04a}} \\
\end{tabular}
\end{center}
\end{table*}

\section{Observations}

Unless otherwise stated the red-end spectra and the photometric
observations were carried out from the 1.3m Telescope of the Skinakas
Observatory (SKI), located in the island of Crete (Greece). Optical
photometry was made using a $1024 \times 1024$ SITe CCD chip with a 24
$\mu$m pixel size (corresponding to $0.5^{\prime\prime}$ on sky). Standard
stars from the \citet{lan92} and \citet{oja96} lists  were
used for the transformation equations.  Reduction of the data was carried
out in the standard way using the IRAF tools for aperture photometry. For
the optical spectroscopic observations the 1.3m telescope was equipped
with a 2000$\times$800 ISA SITe CCD and a 1302 l~mm$^{-1}$ grating, giving
a nominal dispersion of 1.04 \AA/pixel, except for the June 2004 and July 2004
campaigns for which a 1024$\times$1024 CH260 chip (1.3 \AA/pixel) was
employed. The reduction of the spectra was made using the STARLINK {\em
Figaro} package \citep{sho01}, while their analysis was
performed using the STARLINK {\em Dipso} package 
\citep{how98}. For \igrj\ and \igr\ the Skinakas data were complemented
with data from the 2.56-m Nordic Optical Telescope (NOT) and the 2.5-m
Isaac Newton Telescope (INT). 
The results of our photometric and spectroscopic observations are
summarised in Tables \ref{phot} and \ref{spec}.

\begin{table}
\begin{center}
\caption{H$\alpha$ equivalent width measurements}
\label{spec}
\begin{tabular}{cccc}
\noalign{\smallskip} \hline \hline
Date	&MJD			&EW(H$\alpha$)		&Profile\\
	&				&(\AA)			&	\\
\hline 
\multicolumn{4}{c}{\xte} \\
\hline
23-06-03	&53180.44	&--6.3$\pm$0.8  &single peak \\
25-06-04	&53182.40	&--6.8$\pm$0.9  &single peak \\
06-07-04	&53193.38	&--8.2$\pm$1.5  &single peak \\
07-07-04	&53194.37	&--7.3$\pm$1.0  &single peak \\
08-07-04	&53195.37	&--7.5$\pm$1.0  &single peak \\
25-08-04	&53243.27	&--5.5$\pm$0.5  &single peak \\
\hline
\multicolumn{4}{c}{\gro} \\
\hline
25-06-04$^a$	&53182.44	&--4.2$\pm$0.6  &double peak \\
06-07-06$^a$	&53193.38	&--5.0$\pm$0.9  &double peak \\
25-08-04$^a$	&53243.52	&--4.7$\pm$0.5  &double peak \\
13-09-04	&53262.49	&--4.9$\pm$0.5  &double peak \\
03-10-04	&53282.38	&--4.6$\pm$0.6  &double peak \\
24-10-04	&53303.39	&--3.7$\pm$0.4  &double peak \\
\hline
\multicolumn{4}{c}{\sax} \\
\hline
01-08-03$^b$	&52853.50	&--2.2$\pm$0.3  &V$>$R \\
17-08-03$^b$	&52869.53	&--1.1$\pm$0.2  &shell \\
14-09-03$^b$	&52897.36	&+2.3$\pm$0.2	&absorption \\
06-10-03$^b$	&52919.33	&+2.2$\pm$0.3	&absorption \\
08-10-03$^b$	&52921.32	&+2.0$\pm$0.3	 &absorption \\
23-05-04	&53149.56	&+2.6$\pm$0.3	&absorption \\
27-05-04	&53153.57	&+2.5$\pm$0.3	&absorption \\
28-05-04	&53154.52	&+2.4$\pm$0.3	&absorption \\
23-06-04	&53180.52	&+1.8$\pm$0.5	&absorption \\
25-06-04	&53182.47	&+2.2$\pm$0.3	&absorption \\
07-07-04	&53194.54	&+2.7$\pm$0.3	&absorption \\
26-08-04	&53244.44	&+2.1$\pm$0.3	&absorption \\
27-08-04	&53245.44	&+1.6$\pm$0.5	&absorption \\
12-09-04	&53261.36	&+2.1$\pm$0.3	&absorption \\
13-09-04	&53262.35	&+2.1$\pm$0.3	&absorption \\
25-10-04	&53304.35	&+2.1$\pm$0.5	&absorption \\
\hline
\multicolumn{4}{c}{\igrj} \\
\hline
21-05-04	&53147.56	    &+1.1$\pm$0.1    &absorption \\
22-05-04	&53148.55	    &+1.06$\pm$0.05   &absorption \\
24-06-04	&53181.54	    &+1.17$\pm$0.05   &absorption \\
07-07-04	&53194.57	    &+0.81$\pm$0.03   &absorption \\
13-09-04	&53262.39	    &+0.93$\pm$0.03   &absorption \\
24-10-04	&53303.49	    &+0.71$\pm$0.02   &absorption \\
\hline
\multicolumn{4}{c}{\igr} \\
\hline
06-07-04	&53193.53		&52$\pm$2	&single peak \\
26-08-04	&53244.56		&51$\pm$2	&single peak \\
12-09-04	&53261.46		&51$\pm$2	&single peak \\
13-09-04	&53262.40		&53$\pm$4	&single peak \\
03-10-04	&53282.47		&48$\pm$3	&single peak \\
04-10-04	&53283.57		&48$\pm$2	&single peak \\
24-10-04	&53303.58		&49$\pm$3	&single peak \\
\noalign{\smallskip} \hline \hline
\multicolumn{4}{l}{$a$: From \citet{wil05}} \\
\multicolumn{4}{l}{$b$: From \citet{reig04a}} \\
\end{tabular}
\end{center}
\end{table}

\section{Methodology}

In searching for the optical counterparts of new X-ray transients we have
restricted our search to HMXB systems. If the only available information is
that provided by an isolated X-ray detection then we select systems
exhibiting properties of a magnetised neutron star, namely, X-ray
pulsations in the range $1-10^3$ seconds and/or an absorbed power-law
continuum spectrum with an exponential cutoff at 10-30 keV and possible
cyclotron absorption features. Extra information on the nature of the
source can be obtained if the long-term X-ray variability is known
(information provided by all-sky monitors, like RXTE ASM or CGRO BATSE).
The presence of regular and periodic outbursts or unexpected giant
outbursts may indicate the presence of a Be star.

If the distance to the source is known, one can make use of the narrow
spectral range of the optical companion in HMXBs, namely O9-B2, to apply
selection criteria based on the photometric magnitudes and colours. But
without prior knowledge of the distance, which is normally the case, the
most obvious observational feature to look for is the presence of
H$\alpha$ in emission. Note that H$\alpha$ in emission is not only a
property of Be stars but also of many supergiant companions. 

The size of the X-ray error circle determines the type of observational
technique to use. If the X-ray uncertainty region is small (a few
arcseconds) then the number of visible stars in the region is expected to
be small and it is possible to perform narrow-slit
spectroscopic observations and look for early-type, 
H$\alpha$ emitting line stars. If the error radius is large, then it is
likely to include a large number of sources and hence narrow-slit
spectroscopy becomes impractical. Furthermore, uncertainty regions are
given to a certain percentage of probability, what makes it possible that
the true optical companion lies close but outside the X-ray error box.

The size of the X-ray error radius of the sources reported in this work at
the time of the observations was relatively large, typically  $\sim
2^{\prime}$. Therefore we proceeded as follows: the fields around the
best-fit X-ray position were observed through the $B$, $V$, $R$ and $I$
filters and a narrow filter centred at 6563 \AA\ (H$\alpha$ filter). The
instrumental magnitudes corresponding to the $B$, $V$, $R$ and H$\alpha$
filters were used to define a "blue" colour ($B-V$) and a "red" colour
($R-H\alpha$). Then a colour-colour diagram was constructed by plotting
the red colour as a function of the blue colour.  Stars with a moderately
or large H$\alpha$ excess can be distinguished from the rest because they
deviate from the general trend and occupy the upper left parts of the
diagram. Be star are expected to show low $(B-V)$ colours because they are
early-type stars (although they normally appear redder than non-emitting B
stars due to the circumstellar disk) and also larger (i.e., less negative)
$R-H\alpha$ colours because they show H$\alpha$ in emission. This kind of
diagrams have been successfully used to identify optical counterparts in
the Magellanic Clouds \citep{greb97, ste99}.

Good candidates are then those that occupy the upper left parts of the
colour-colour diagram {\em and} lie inside or very close to the X-ray
uncertainty region. Under these conditions the colour-colour diagram
restricts the possible candidates to a handful of sources (sometimes just
one or two), hence making narrow-slit spectroscopy useful. The next step
is to obtain H$\alpha$ spectra of those potential candidates. In order to
check for the usefulness of the diagram and secure our method we also
obtained spectra of the brightest stars inside the X-ray error circles.
With the appropriate instrumental set-up the red-end spectrum can extend
beyond 6700 \AA, hence the \ion{He}{i} line at 6678 \AA, if present, can
be taken as evidence that the optical counterpart is indeed an early-type
object. Nevertheless, a classification spectrum (i.e. around 4000-5000
\AA) is desirable to pin down the exact spectral type.

Figure \ref{cdimag} shows the colour-colour diagrams and the optical
images corresponding to the visual band of the field around the X-ray
positions of \xte\ and \igr. The optical counterparts have been marked
with a filled circle on the colour-colour diagrams and with an arrow on
the images. The satellite uncertainty circles at the time when the optical
observations were performed are also indicated.

In order to assess the statistical significance of this method we can
estimate the expected number of Be/X-ray binaries present in a circular
portion of the sky with 2$^{\prime}$ radius (i.e., the typical size of the
satellite uncertainties). Assuming that there are about 0.005-0.5 Be stars
per arcmin$^2$ in the Galactic Plane and that about $2/9$ of the Be stars
take part in Be/X-ray binaries \citep{wil05}, we estimate the total number
of Be/X-ray binaries inside the typical error circles considered in this
work in $\sim 0.01-1$.  Given that Be/X-ray binaries represent more than
70\% of the HMXBs and the uncertainty in the above calculation, we can
assume that the range 0.01--1 is also representative of all types of
HMXBs.

\begin{figure}
\begin{center}
\includegraphics[width=8cm,height=6cm]{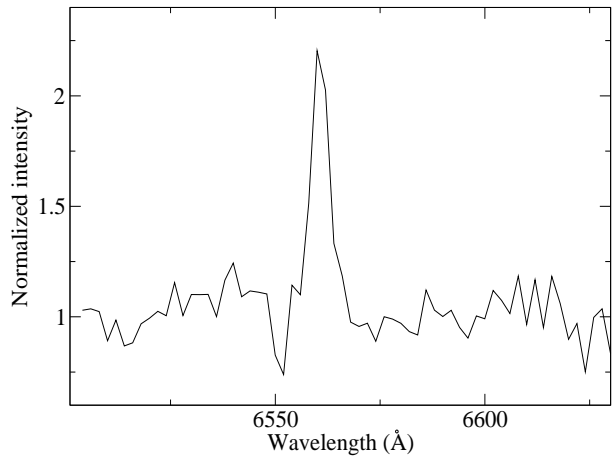} 
\caption[]{Average H$\alpha$ profile of \xte.
The spectrum was rebinned with a bin size of 2 \AA.}
\label{xte1858_pcyg}
\end{center}
\end{figure}

\section{Results and discussion of individual sources}

\subsection{\xte}

\xte\ was discovered by the RXTE satellite in 1998 with an estimated
uncertainty (90\% confidence) of 6$^\prime$. The peak flux, 24 mCrab (2-12
keV), was reached on 7 February 1998 \citep{rem98}. The X-ray position was
later refined down to a 90\% confidence error radius of
$2^{\prime}.5$ \citep{mar98}. X-ray pulsations with a pulse period of
221.0$\pm$0.5 s \citep{tak98} and quasi-periodic oscillations centred at
0.11Hz \citep{pau98} were also detected. The pulsations and transient
nature of \xte\ suggested that it is a Be/neutron star binary. 
\citet{mol04} reported the detection of a new X-ray outburst in April 2004
with INTEGRAL and an improvement of its coordinates. The average flux was
$\sim$ 20 mCrab and $\sim$15 mCrab in the 3-15 keV and 18-60 keV energy
bands, respectively. The uncertainty of the INTEGRAL IBIS and JEM-X
detectors was 2$^\prime$ and $1^\prime$, respectively.

Optical photometric observations of the field around the best-fit INTEGRAL
position were carried out on the nights 20 May 2004, 27 July 2004 and 24
August 2004. Based on positional coincidence several
candidates were identified (see Fig.~\ref{cdimag}). Subsequent
spectroscopic observations revealed that only one exhibited H$\alpha$
emission. The position of the proposed candidate is R.A.= 18h58m36s,
Decl.=03d26m09s, consistent with the ISGRI/IBIS uncertainty, although
outside the JEM-X error circle. No photometric or spectroscopic
variability has been detected during the three months covered by our
observations (Table~\ref{phot} and \ref{spec}). Unfortunately, the
weakness of the source in the blue band prevented us from obtaining a good
enough S/N spectrum, hence, no spectral classification analysis has been
made so far. Although the identification of \xte\ with a HMXB is apparent
we cannot tell the spectra subtype nor the luminosity class. Some spectra
seem to show a trough bluewards of the central peak, reminiscent of a P-Cygni
profile and possibly indicating a supergiant companion
(Fig.~\ref{xte1858_pcyg}). The relatively long spin period (221 s) would
give further support to the presence of an evolved companion as transient
BeX systems tend to have shorter spin periods ($P_{\rm spin} \simless 100$
s). However, the strength of the H$\alpha$ line ($\sim -7$ \AA) is somehow
larger than average values found in SXRBs ($\simless$ 5 \AA). Strong
H$\alpha$ emission and P-Cygni profiles has been reported for the
hypergiant system \object{Wray977}/\object{2S 1223-624} \citep{kap95}.
Blue-end spectra and higher quality red-end spectra are needed to determine
the nature of \xte.

\begin{figure}
\begin{center}
\includegraphics[width=8cm,height=6cm]{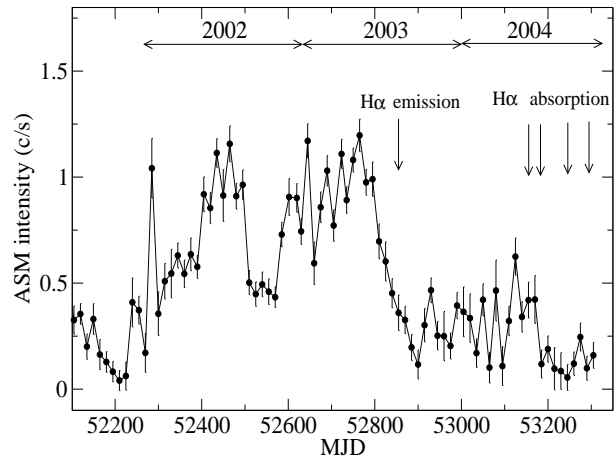} 
\caption[]{X-ray/optical evolution of \sax. Absorption H$\alpha$ profiles
correspond with low X-ray activity states.}
\label{sax2103_xopt}
\end{center}
\end{figure}

\subsection{\gro}

\gro\ was discovered in September 1995 by GRO BATSE during a giant
outburst that lasted for 46 days \citep{wil98}. The total
flux reached about 300 mCrab in the energy range 20-50 keV. Based on the
regularity of a series of smaller (15-20 mCrab) outbursts following the
first giant outburst the orbital period was proposed to be 110 days. BATSE
also detected shorter and weaker outbursts halfway between the intermediate
outbursts, which opened the possibility that the orbital period was 55 days.
Recent timing analysis with RXTE ASM data favour the value of 55 days as the
correct orbital period \citep{wil05}.

The uncertainty in the location of \gro\ given by BATSE was initially a
95\% confidence box of $4^{\circ} \times 1^{\circ}$, which was improved by
OSSE to $30^{\prime} \times 60^{\prime}$ (95\% confidence). The X-ray
position was further refined by RXTE PCA to a radius error circle of
$4^{\prime}$ (90\% confidence). With this uncertainty in the finding
charts we performed our optical observations resulting in the discovery of
the optical counterpart \citep{reig04b}. A recent {\it
Chandra} observation pinned down the position of \gro\ to just
$0^{\prime \prime}.41$ (68\% confidence radius), consistent with our
optical position \citep{wil05}. 

The optical counterpart to \gro\ has been identified with a V=14.9
O9.5-B0IV-Ve star located at a distance of $\sim 9\pm1$ kpc.  After a
period of high X-ray activity (1995-2002) \gro\ entered a quiescent phase
in mid 2002 that continues up to the present. The {\it Chandra}
observations of February 2004 failed to detect pulsations and estimated
the unabsorbed 2--10 keV flux to be $(3-9) \times 10^{33}$ erg s$^{-1}$.
No optical variability is apparent during the four months (June-October
2004) that we monitored the source. The photometric magnitudes remained
unchanged at mean values of $B=16.05$, $V=14.93$, $R=14.22$ and
$I=13.44$. 

Despite the fact that the X-ray observations showed a quiescent state, the
Be star did not lose the disk completely. The H$\alpha$ line exhibited
always a double-peak emission profile with a mean equivalent width of
$-4.7$ \AA\ and a peak separation of 400$\pm$30 km s$^{-1}$. The presence
of the disk can also explain the higher scatter of the photometric
magnitudes as the wavelength increases. The standard deviation of the
observations of Table~\ref{phot} for the various photometric bands are:
$\sigma_B=0.018$, $\sigma_V=0.023$, $\sigma_R=0.030$ and $\sigma_I=0.050$.

\subsection{\sax}

The X-ray transient \sax\ was discovered by the BeppoSAX Wide Field Camera
(WFC) during an outburst in February 1997, reaching a peak intensity of 20
mCrab (2--25 keV). The source was found to be an X-ray pulsar with a
358.61 second pulse period \citep{hul98}. Since its discovery \sax\ has
been detected by other three satellites: RXTE observations allow the
determination of the orbital parameters: $P_{\rm orb}=12.68$ d,
$e=0.4\pm0.2$ \citep{bay00}; XMM-Newton observations detected
quasi-periodic oscillations at 0.044 Hz and estimated the magnetic field
of the neutron star to be $\sim 7\times 10^{12}$ G \citep{ina04}; INTEGRAL
observations provided the first broad-band spectrum and showed significant
emission up to 150 keV \citep{bla04}.

Optical observations reporting on the identification of the optical
counterpart with a V=14.2 B0Ve star at 6.5 kpc has been reported elsewhere
\citep{reig04a}. Here we present new photometric and spectroscopic
observations performed in 2004. At the time of the optical identification
(mid 2003) the source was caught in the final stages of a disk-loss
phase.  All new observations show H$\alpha$ in absorption, indicating that
the Be star has not recovered the circumstellar disk. The source is known
to go through high and low X-ray states \citep{bay00}. Although the
optical coverage is still scarce, a loose correlation
between the X-ray and optical data is present. As can be seen in
Fig.\ref{sax2103_xopt}, where a portion of the RXTE ASM of \sax\ is
plotted, the emission profiles were observed at the end of a high X-ray
state, whereas the absorption profiles correspond to a low activity state
which \sax\ entered at the end of 2003 and still continues. This is quite
a puzzling situation since without the disk no X-ray emission should be
detected --- the material in the disk constitutes the fuel that powers the
X rays. However, the source is still active in X-rays as can be noticed
from the RXTE ASM. We discuss this issue further in Sect.~\ref{quies}.

\subsection{\igrj}

The X-ray source IGR J00370+6122 was detected by {\it INTEGRAL} during
December 2003. The error circle ($\sim$~2$^\prime$) included the position of
the {\it ROSAT} source 1RXS J003709.6+612131 \citep{har04}, which had
been identified with the OB star BD~$+60\degr$73 = LS~I~$+61\degr$161
\citep{rut00}. A search of {\it RossiXTE}/ASM data from this position
revealed the presence of a variable X-ray source, displaying a clear
modulation with a period of $15.665\pm0.006\:$d. When the light curve is
folded on this period, the source is only clearly detected as a single
peak close to the phase of maximum \citep{har04}.

\begin{figure}
\begin{center}
\includegraphics[width=6cm,height=8cm,angle=-90]{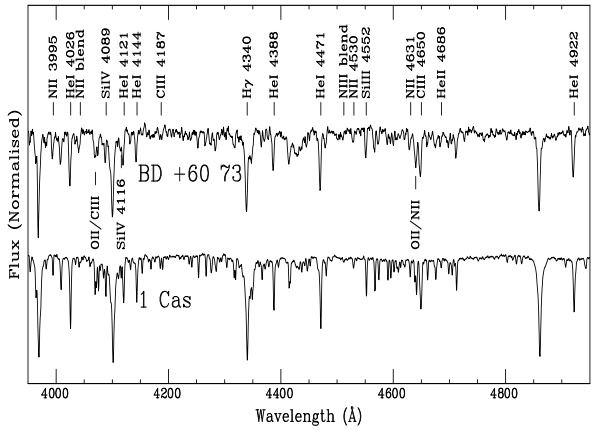} 
\caption[]{Blue-end spectrum of \igrj. A comparison with a B0.5III standard
is provided.}
\label{igr0037_sp}
\end{center}
\end{figure}

We performed five spectroscopic and two photometric runs from Skinakas
observatory during the summer 2004 (see Table~\ref{phot} and \ref{spec}).
Additionally, an intermediate-resolution spectrum of this object was taken
on the night of 2003 July 7th with the 2.5-m Isaac Newton Telescope (INT),
at La Palma, Spain. The INT was equipped with the Intermediate Dispersion
Spectrograph (IDS) and the 235-mm camera. The R900V grating and EEV\#13
camera were used, resulting in a nominal dispersion of
$\approx0.65$\AA/pixel.  This spectrum is shown in Fig.~\ref{igr0037_sp},
together with that of HD 218376 (1~Cas), given as B0.5\,III standard by
\citet{wal71}. In both stars, the only \ion{He}{ii} line visible is
4686\AA, implying a spectral type earlier than B1 but later than B0.2
\citep{wal90}. BD~$+60\degr$73 was classified as B1Ib by \citet{mor55}.

Comparison of the two spectra clearly shows BD~$+60\degr$73 to be more
luminous than 1~Cas (the Balmer lines are less deep and have narrower
wings) in spite of the higher rotational velocity indicated by the broader
\ion{He}{i} and metallic lines. However, the moderate strength of
\ion{Si}{iv}~4089\AA\ and \ion{Si}{iii}~4552\AA\ prevents a supergiant
classification. The most adequate classification appears to be
B0.5\,II-III, where the composite luminosity class indicates an
intermediate luminosity. Though BD~$+60\degr$73 appears slightly earlier
than 1~Cas (as it displays stronger \ion{He}{ii}), a spectral type B0.2
cannot be given because of the absence of the \ion{He}{ii}~4541\AA\ line.
The low-resolution spectra covering the yellow/red region of the spectrum
obtained from the Skinakas Observatory show that \ion{He}{i}~5875\AA,
\ion{He}{i}~6678\AA\ and H$\alpha$ are all strongly in absorption, as is
morphologically normal for the spectral type. BD~$+60\degr$73 has thus not
entered a Be phase. Another interesting different between the spectrum of
BD +60 73 and that of the standard is the greater strength of all the 
\ion{N}{ii} and \ion{N}{iii} lines in the former.
This N enhancement does not seem to be accompanied by any
C depletion, perhaps suggesting the accretion of CNO processed material
during a previous evolutionary phase, when the progenitor of the compact
object was a giant or supergiant. 

$UBV$ photometry of BD~$+60\degr$73 is reported by \citet{hil56} and
\cite{hau70}. Their values are identical within the errors (the photometry
of Haug is transformed to Hiltner's system) and give $V=9.64$,
$(B-V)=0.57$, in complete agreement with our photometric measurements
(Table~\ref{phot}). Assuming the intrinsic colours of a B0.5 giant
\citep{weg94} and $M_{V}=-5.2$, intermediate between the values for
luminosity class II and III, we obtain $E(B-V)=0.78$ and $DM=12.4$,
corresponding to a distance $d=3.0\:$kpc. Such values are typical of
Perseus Arm objects in the Cassiopeia region and suggest that
BD~$+60\degr$73 may belong to the "far" component of Cas~OB5 
\citep[cf.][]{amp64}.

Because of the spectral type of its counterpart, \igrj\ appears difficult
to fit within the classification scheme of massive-X-ray binaries. The
counterpart is neither a Be star nor a supergiant. It is not surrounded by
a circumstellar disk and it is expected to have only a moderate wind,
rising the question of how the X-ray emission originates. The fact that
the X-ray light curve seems to be very strongly peaked supports the idea
of a very eccentric orbit. If this is the case, the compact object could
come very close to the B-type star at periastron, allowing accretion from
the inner regions of its wind, or perhaps triggering some extra mass loss.
In this sense, \igrj\ shares some similarity with the extreme LMC X-ray
transient \object{A\,0538$-$66} \citep{cha83}, though obviously the X-ray
activity is much lower. Depending on the orbital parameters, \igrj\ could
evolve toward a Roche-lobe overflow system such as \object{LMC X-4} (if
the orbit circularises on a time-scale smaller than the evolutionary time
for the mass donor) or, more likely, a SXB, if the mass donor swells to
supergiant size while the orbit is still relatively broad. Though no
pulsations have been detected from \igrj, the lack of radio emission
\citep{cam04} hints at the presence of a neutron star in this system.

\begin{figure}
\begin{center}
\includegraphics[width=8cm,height=6cm]{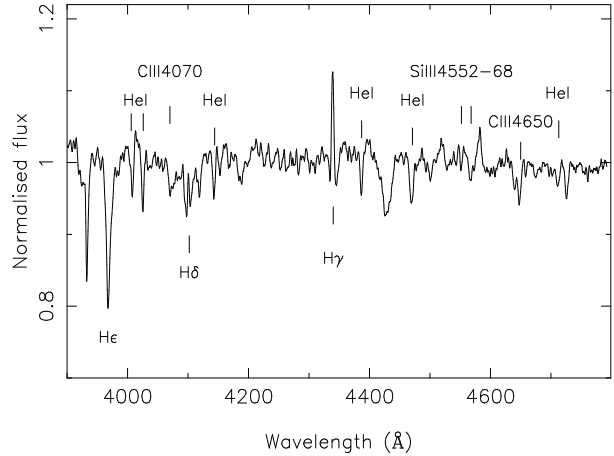} 
\caption[]{Blue-end spectrum of \igr. The He I lines are 4006 \AA, 4026
\AA, 4144 \AA, 4387 \AA, 4471 \AA\ and 4713 \AA}
\label{igr0136_sp}
\end{center}
\end{figure}

\subsection{\igr}

\igr\ was discovered by INTEGRAL imager IBIS/ISGRI on April 19, 2004
during observations dedicated to the Galactic Plane Scan. The source
position was provided with an uncertainty $2^\prime$. The average
flux over 2.3 hours of observations was 17 mCrab in the 17-45 keV band.
The source was not detected with IBIS/ISGRI during a Galactic Plane Scan
of the region on May 1, 2004 \citep{gre04}. 

Our optical observations revealed that the star located at  R.A.=
01h35m50s, Decl.=66d12m40s, consistent with the INTEGRAL  uncertainty
circle (see Fig.~\ref{cdimag}), shows a very strong H$\alpha$ emission
\citep{reig04c}. Other potential candidates (star 1 in
Fig.~\ref{cdimag}) showed H$\alpha$ in absorption.

In addition to the Skinakas observations \igr\ was also observed with the
Nordic Optical Telescope on 2004 October 4th using the
Andalucia Faint Object Spectrograph and Camera (ALFOSC). The instrument
was equipped with a thinned $2048\times2048$ pixel E2V CCD. Observations
of the field of \igr\ were taken in slitless spectroscopy mode using
the Bessell R-band filter and grism \#4. Spectra of the candidate were
taken with grisms \#16 and \#7 which cover the 3500--5100\AA\ and
3800--6870\AA\ range respectively, with nominal dispersions of
0.8\AA/pixel and 1.5\AA/pixel respectively.

The classification spectrum (4000-5000 \AA) shows the presence of H$\beta$
and H$\gamma$ in emission and the metallic and \ion{He}{i} lines typical
of an early B-type star of low luminosity (Fig.\ref{igr0136_sp}). More
precisely, the absence of \ion{He}{ii} lines indicates an spectral type
later than B0.5.  \ion{He}{i} lines dominate the spectrum (together with
the hydrogen lines),  indicating a B1 or B2 star. However, the presence of
some amount of \ion{Si}{iii}~4552-68\AA\ and the carbon blends
(\ion{C}{iii}~4070--4650\AA) favours the B1 spectral type. On the other
hand, the weakness of the oxygen and silicon lines points toward a
main-sequence star, although the strength of \ion{C}{iii}~4650\AA\ is more
typical of giant companion. Thus we conclude that the optical counterpart
to \igr\ is a B1V star, although a more luminous companion cannot be ruled
out. This object is probably identical with the catalogued emission line
star [KW97] 6-30 \citep{koh99}, though the coordinates listed in this
catalogue locate the source about $2^{\prime}$ south-west of our proposed
candidate and outside the INTEGRAL error circle. \igr\ shows the second
strongest, after \object{A 1118-616}, H$\alpha$ emission among galactic
BeX with an equivalent width of $\sim$ 50 \AA.

Measured photometric magnitudes are given in Table~\ref{phot}.  Taking the
typical colour of a B1V star, $(B-V)_0=-0.23$ \citep{weg94}, and
comparing with the observed colour  $(B-V)_{\rm obs}=1.38$, we derived an
excess colour of $E(B-V)=1.61$. Assuming an absolute magnitude of
$M_V=-3.2$ \citep{jas87}, we estimate the distance to be $\sim$2 kpc,
compatible with the near side of the Perseus Arm. 

\begin{table*} 
\begin{center} 
\caption{Optical and X-ray properties of the
sources} 
\label{prop} 
\begin{tabular}{lccccc|ccc} 
\noalign{\smallskip} \hline \hline \noalign{\smallskip}
\multicolumn{6}{c}{Optical data} &\multicolumn{3}{c}{X-ray data} \\ 
Source	&R.A.  &Dec. 	&spectral	&E(B-V) &Distance &P$_{\rm spin}$	&QPO	&Peak flux (mCrab)/\\ 
	&(2000)	&(2000)		&type		&	&(kpc)	 &(s)			&(Hz)	&Energy range \\ 
\noalign{\smallskip} \hline  \noalign{\smallskip}
\xte	&18 58 36&03 26 10	&B?e		& &	&221	&0.11	&24 (2-12 keV) \\ 
\gro	&20 58 47&41 46 36	&O9.5-B0IV-Ve	&1.40	&9.0	&192	&	&300 (20-50 keV) \\
\sax	&21 03 36&45 45 04	&B0Ve		&1.35	&6.5	&358	&0.044	&27 (2-12 keV)\\ 
\igrj	&00 37 10&61 21 35	&BN0.5II-III	&0.75	&3.3	&	&	&15 (2-28 keV)	\\ 
\igr	&01 35 50&66 12 40	&B1IV-Ve	&1.61	&2.0	&	&	&19 (17-45 keV)
\\	 
\noalign{\smallskip} \hline
\end{tabular} 
\end{center} 
\end{table*}

\section{BeX-ray binaries in quiescence}
\label{quies}

The study of BeX in quiescence is still an unexplored field.
So far only a handful of BeX systems have been observed in quiescence: 
\object{A0538-66}, \object{4U 0115+63}, \object{V0332+53} \citep{cam02},
\object{A0535+262} \citep{neg00} and \gro\ \citep{wil05}.

In the X-ray band the limited number of observations \citep{cam02,wil05}
show very soft spectra, absence of pulsations (except in
A0535+262) and luminosities well below $10^{34}$ erg s$^{-1}$.  Quiescent
states are generally related to the propeller effect \citep{ill75}. There
exists a minimum X-ray luminosity below which the propeller effect sets
in. This luminosity is given by \citep[see e.g][]{cam02}

\begin{eqnarray}
L_{\rm min}(R_{\rm NS}) &=& 3.9 \times 10^{37} 
\left(\frac{B}{10^{12} {\rm \, G}}\right)^2
\left(\frac{P_{\rm spin}}{1 \, {\rm s}}\right)^{-7/3} \nonumber \\
&& \left(\frac{M_{\rm X}}{1.4 \, {\rm \msun}}\right)^{-2/3}
\left(\frac{R_{\rm X}}{10^6 \, {\rm cm}}\right)^5 \, \,  {\rm erg \,
s^{-1}} \nonumber
\end{eqnarray}

\noindent Below this limiting luminosity the accreting matter can no
longer reach the neutron star surface because it is spun away by the
fast rotation of the magnetosphere. For rapid rotating pulsars like 4U
0115+63 or V0332+53 $L_{\rm min}(R_{\rm NS})$ is relatively high,
$\sim 10^{36}$ erg s$^{-1}$. For long pulsing systems like \gro, \sax\ and
A0535+262, $L_{\rm min}(R_{\rm NS}) \sim 10^{32}-10^{33}$ erg s$^{-1}$.
When the propeller effect is at work, X-rays can still be produced through
accretion onto the magnetosphere \citep{cor96}, as opposed to onto the
neutron star surface, or through leakage through the magnetosphere
\citep{cam01}. A third model that has been put forward to explain X-ray
emission during quiescence, namely, cooling of the neutron star surface after
crustal heating \citep{bro98}, can be ruled out  as the general mechanism
since in order to achieve the observed luminosities it requires prior
events of intense accretion, that is, a high outbursting activity before
entering the quiescent state, which was not seen in A0538-66, V0332+53 or
A0535+262. It, however, might still contribute in part to the total
quiescent luminosity.

In this work we focus on the optical state of BeX during X-ray quiescence.
In the standard model of BeX systems, the circumstellar disk around the Be
star's equator provides the reservoir of matter that ultimately is
accreted onto the neutron star. Thus, one would expect that during the
X-ray quiescent state the disk would be missing or largely debilitated.
This seems to be the case for \gro, for which  optical observations during
the X-ray quiescent state showed H$\alpha$ in emission, i.e, the
disk was still present. Given the relatively low $L_{\rm min}(R_{\rm NS})$
for \gro, mass transfer may still be going on during quiescence states
from a weak disk and there is no need to invoke thermal emission from
crustal heating or accretion onto the magnetosphere.

The cases of \sax\ and \object{A0535+262} are particularly interesting
because despite the complete loss of the circumstellar disk X-ray 
emission was detected 
\citep[for a detailed discussion of \object{A0535+262} see][]{neg00}.  
All optical spectra of \sax\ taken during 2004 exhibit
H$\alpha$ in absorption. Still, the source shows X-ray activity with a
mean flux of $\sim 3-5 \times 10^{35}$ erg s$^{-1}$ ($\sim 0.25$ ASM c
s$^{-1}$), as can be seen in Fig.~\ref{sax2103_xopt}. We assume a distance
of 6.5$\pm$0.9 kpc \citep{reig04a}. This luminosity is above $L_{\rm
min}(R_{\rm NS})$, hence accretion is likely to be the mechanism that
powers the X-rays but contrary to \gro, \sax\ has completely lost the disk.

\sax\ is the Galactic BeX with the narrowest orbit, $P_{\rm orb}=12.6$
days, and shows moderately eccentricity, $e \approx 0.4$ \citep{bay00}.
Interestingly, \sax\ occupies the region of the wind-fed supergiant binaries
in the $P_{\rm spin}-P_{\rm orb}$ diagram \citep{reig04a}. In principle,
then, accretion from the stellar wind of the B0 companion might be at the
origin of the observed luminosity, as much the same way as we proposed to
be the case in \igrj, for which $L_{\rm X} (1.5-12 \, {\rm keV}) \approx
(0.5-2) \times 10^{35}$ erg s$^{-1}$ \citep{har04}. We can make an
estimate in order of magnitude of the expected luminosity using a simple
wind-fed model. Assuming that all the gravitational energy is converted
into X-rays, the X-ray luminosity is given by \citep[see e.g][]{wat89}

\begin{eqnarray}
L_{\rm X} &\approx& 4.8 \times 10^{37} 
\left(\frac{M_{\rm X}}{1.4 {\rm \msun}}\right)^3 
\left(\frac{R_{\rm X}}{10^6 \,{\rm cm}}\right)^{-1}
\left(\frac{M_*}{1 {\rm \msun}}\right)^{-2/3}  \nonumber \\
&&\left(\frac{P_{\rm orb}}{1 \, {\rm day}}\right)^{-4/3}  
\left(\frac{\dot{M}_*}{10^{-6} {\rm \msun yr^{-1}}}\right) 
\left(\frac{v_{\rm w}}{10^8 \,{\rm cm \, s^{-1}}}\right)^{-4} 
\, \, {\rm erg \, s^{-1}} \nonumber
\end{eqnarray}

Taking $M_*=20$ $\msun$ \citep{vac96}, the canonical mass and radius for
a neutron star, $M_{\rm X}=1.4$ $\msun$ and $R_{\rm X}=10^6$ cm, the
corresponding orbital periods of 12.68 d for \sax\ and 
15.66 d for \igrj, and wind velocities in the range 400-1000 km s$^{-1}$ 
\citep{sno81,pri89} we note that mass-loss rates of the order
$10^{-7}$ $\msun$ yr$^{-1}$ are needed to reproduce the observed X-ray
luminosity. This mass-loss rate is not unusual for early-type giants and
supergiants \citep{pri90}. 

For \sax\ one has to invoke the asymmetry of mass outflows that
characterise Be stars if the observed X-ray luminosity is to be explained
by wind accretion. The structure of the stellar winds in Be stars is
highly asymmetric and contains two components: at higher latitudes, mass
is lost through the high-velocity, low-density wind; in the equatorial
regions a slower and denser wind operates and ultimately forms the disk.
Equatorial mass-loss rates are of the order of $10^{-7}$ $\msun$ yr$^{-1}$
and are a factor 10-100 larger than polar mass-loss rates 
\citep[][and references therein]{wat88}.  
Note that in the case of \sax\ there is no
need to reach $\dot{M} \sim 10^{-7}$ $\msun$ yr$^{-1}$ due to its narrow
orbit and the strong dependence of the luminosity on the wind velocity
($L_{\rm X} \propto v_{\rm w}^{-4}$). Stellar winds are generally modelled
with a wind velocity field of the form \citep{cas75}

\[  v_{\rm w}(r)=v_{\infty}\left(1-\frac{R_*}{r}\right)^{\beta}   \]

\noindent where $\beta$ is normally taken to vary in the range 0.7--1 and
$R_*$ is the star radius (8.3 $\rsun$ for a B0). The
relatively narrow orbit of \sax\ implies that the wind does not reach its
terminal velocity, $v_{\infty}$, but varies in the range $v_{\rm w} \sim
[0.7-0.9]v_{\infty}$.

\section{Summary and conclusions}

We have presented the result of our search for optical counterparts to
high-mass X-ray transients. We have used a method based on the combination
of photometric colour-colour diagrams with medium-resolution spectroscopy.
This method rests on two observational properties of this type of systems,
namely, the narrow range of spectral types and the fact that the vast
majority of the optical companions in high-mass X-ray binaries show
H$\alpha$ in emission. This method has been proved to be very  useful  in
cases where the uncertainty  in the position of the X-ray source is large
(a few arcminutes). We found the optical counterparts to \xte, \gro, \sax\
and \igr.  Three systems, \gro, \sax\ and \igr\ harbour main-sequence or
subgiant B stars and hence constitute new BeX binaries. The nature of
\xte\ is not clear:  its H$\alpha$ line resembles a  P-Cygni profile
reminiscent of very luminous stars with strong stellar winds but its
transient nature is more typical of BeX systems. If it contains an evolved
companion, then the large equivalent width might be indicative of an
hypergiant star. We have carried out a detailed spectroscopic analysis of
\igrj\ and \igr\ and found that the systems contain an evolved BN0.5
II-III and a main-sequence B1 companion, respectively. The production of
X-rays in \igrj\ is difficult to explain unless a very eccentric orbit is
invoked. Determination of the orbital parameters of \igrj\ could provide
important clues on the evolutionary path leading to the formation of
high-mass X-ray binaries. Finally, a study of the correlated X/optical
data of \sax\ and \gro\ during quiescence shows that while quiescent
emission in \gro\ can still proceed via accretion from the circumstellar
disk onto the surface of the neutron star, in \sax\ it must come from the
stellar wind. 

With this new findings we have increased the number of known
traditional  HMXBs. However, not all the new suspected HMXBs that are 
being discovered fit into the known subgroups of the known population of 
massive binaries. Since its launch, and mainly thanks to its continual
monitoring of the Galactic plane, {\it INTEGRAL} has detected a large
number of new X-ray sources, most of which are HMXBs
\citep[][]{luto05,kuul05}. Most of them seem to have interesting new
properties, which set them aside from previously known HMXBs. A few
sources display very high variable absorption, indicative of a very dense
local environment. Among them, IGR J16318$-$4848 has been identified with
a very peculiar IR counterpart by \citet{fill04}, who suggest that it may
be a B[e] supergiant. Another subset of sources, less obscured, appear to
be transients with very short outbursts. Three of them are associated with
OB supergiants \citep{smit04,negu05}, and hence are likely to belong to a
new class of HMXB transients.

Other new {\it INTEGRAL} sources seem to belong to well established
classes of HMXB. This is the case of IGR~J01363+6610 or IGR~J11435--6109
\citep{torr04}, which appear to be typical Be/X-ray transients. This last
source, like IGR~J00370+6122, had already been observed by other
satellites, though not recognised as a HMXB. In this sense, it is unlikely
to be coincidental that obscured sources belonging to new classes are
mainly being found in the areas covered by the Galactic Centre scans,
which are deeper and more frequent. This suggests that deep continuous
monitoring of other regions of the Galactic disk would result in the
discovery of many new HMXBs. 

\begin{acknowledgements}

 IN is a researcher of the
programme {\em Ram\'on y Cajal}, funded by the Spanish Ministerio de
Educaci\'on y Ciencia and the University of Alicante, with partial support
from the Generalitat Valenciana and the European Regional Development Fund
(ERDF/FEDER). This research is partially supported by the MEC through
grant ESP-2002-04124-C03-03.
Skinakas Observatory is a
collaborative project of the University of Crete, the Foundation for
Research and Technology-Hellas and the Max-Planck-Institut f\"ur
Extraterrestrische Physik. 
The Nordic Optical Telescope is operated on the island of La Palma
jointly by Denmark, Finland, Iceland, Norway, and Sweden, in the
Spanish Observatorio del Roque de los Muchachos of the
Instituto de Astrofisica de Canarias. Part of the data presented here have
been taken using ALFOSC, which is owned by the Instituto de
Astrof\'{\i}sica
de Andaluc\'{\i}a (IAA) and operated at the Nordic Optical Telescope under
agreement between IAA and the NBIfAFG of the Astronomical
Observatory of Copenhagen.
The INT is operated on the island of La
Palma by the Isaac Newton Group in the Spanish Observatorio del Roque
de Los Muchachos of the Instituto de Astrof\'{\i}sica de Canarias.
Some of the WHT spectra were obtained as part
of the ING service programme. This research has made use of NASA's
Astrophysics Data System Bibliographic Services.

\end{acknowledgements}

\end{document}